\documentclass[aps,pra,twocolumn,superscriptaddress,groupaddress,showpacs,floatfix,notitlepage,nofootinbib]{revtex4-2}

\usepackage{times,graphics,graphicx,color,xcolor,amsfonts,amssymb,amsmath,bm,bbm,dsfont,hyperref}

\hypersetup{colorlinks,linkcolor={blue},citecolor={blue},urlcolor={blue}}
\newcommand{\ignore}[1]{}

\DeclareFontFamily{OT1}{pzc}{}
\DeclareFontShape{OT1}{pzc}{m}{it}%
              {<-> s * [1.25] pzcmi7t}{}
\DeclareMathAlphabet{\mathpzc}{OT1}{pzc}%
                                 {m}{it}
\usepackage[T1]{fontenc} 
\usepackage{newtxtext,newtxmath} 

\begin{document}

\title{Fidelity-based supervised and unsupervised learning for binary classification of quantum states}

\author{F. Shahi}
\affiliation{Department of Physics, Sharif University of Technology, Tehran 14588, Iran}

\author{A. T. Rezakhani}
\email{rezakhani@sharif.edu}
\affiliation{Department of Physics, Sharif University of Technology, Tehran 14588, Iran}
\affiliation{School of Physics, Institute for Research in Fundamental Sciences (IPM), Tehran 19538, Iran}

\begin{abstract}
Here we develop two quantum-computational schemes for supervised and unsupervised classification tasks in a quantum world by employing the quantum information-geometric tools of quantum fidelity and quantum search algorithm. Presuming that pure states of a set of given quantum systems (or objects) belong to one of two known classes, the objective here is to decide to which of these classes each system belongs---without knowing its state. The supervised binary classification algorithm is based on having a training sample of quantum systems whose class memberships are already known. The unsupervised binary classification algorithm, however, uses a quantum oracle which knows the class membership of the states of the computational basis. Both algorithms require the ability to evaluate the fidelity between states of the quantum systems with unknown states, for which here we also develop a general scheme. 
\end{abstract}

\maketitle

\section{Introduction}
\label{sec:int}

Statistical classification is an important goal in modern science and technology \cite{book:Hastie,book:Alpaydin}. One of the simplest examples of such tasks is to decide to which of a pair of specified target groups a particular object of interest is assigned. In general, such tasks can be regarded as trying to figure out if a statement $S_{\mathbf{O}}$ referring to a particular object $\mathbf{O}$ is ``\textsc{true}'' or ``\textsc{false}'' (or ``\textsc{yes}'' and ``\textsc{no},'' alternatively). Any such method is referred to as a ``binary classification'' problem in the context of ``pattern recognition'' \cite{book:Hastie}. Distinguishing between spam or non-spam emails, determining whether a collection of physical symptoms of a patient is due to an underlying disease or not, and identity verification, along with a diverse set of other relevant applications, are all examples of the problem of binary classification. Binary classification is a special case of the problems where the number of target groups is not necessarily two and can be any natural number (or even cardinality of continuum), e.g., handwriting recognition \cite{Xu}.

The capability of most classification methods in assigning labels from a set of target groups to specific types of objects mainly relies on the associate ``training set,'' a relatively small subset of objects having their labels known somehow in advance. Such classification methods in which training sets are used to find hidden structures are referred to as ``supervised machine learning'' algorithms (vs. \textit{unsupervised} machine learning) \cite{book:Hastie}, e.g., ``support vector machine'' \cite{Suykens}. 

Each object $\mathbf{O}$ in machine learning is usually represented using a finite sequence of real numbers that can be regarded as a vector $\mathbf{v}_{\mathbf{O}}$ in a Euclidean space. Naturally then, resemblance between objects is assessed using the distance of their associated vectors. Performing various steps of a machine leaning algorithm would require computational and processing power, which would increase with complexity of algorithms. With the advent of quantum computation, quantum information processing, and quantum algorithms \cite{book:Nielsen}, \textit{quantum} machine learning has also been studied \cite{book:QML}. Most recent attempts in this line have concerned revisiting earlier (classical) machine learning algorithms such that one can best employ power of quantum mechanics \cite{Adcock,Schuld,survey}. For example, recently a quantum version of the support vector machine algorithm has been developed \cite{Lloyd:PRL}, which in turn relies on a quantum algorithm for solving linear systems of equations \cite{Lloyd:linear}.

Quantum machine learning can also be categorized into the classes of supervised and unsupervised quantum classification problems. In the case where the quantum states describing the quantum systems of interest are \textit{known}, the problem of supervised quantum classification almost perfectly resembles its classical counterpart, except for the manner of representing the objects. This difference, however, is not fundamental since there always exists a one-to-one correspondence between the representations of quantum systems in a Hilbert space and vectors of a Euclidean space of the same dimension. Thus, assuming to be equipped with a quantum-\textsc{ram} \cite{RAM}, any scheme for dividing quantum systems into categories according to their known quantum states is also a tool for revealing patterns in classical data. Quantum-\textsc{ram} is defined to be a device which is able to provide a mapping from classical vectors in a desired Euclidean space onto the quantum states living in the corresponding Hilbert space \cite{RAM}.
Another example of this criterion, concerned with classifying two-qubit quantum states, can be found in Ref.~\cite{Guta}. Quantum machine learning can also be extended to devise algorithms for recognizing patterns in quantum systems with \textit{unknown} quantum states \cite{Sentis}. Such algorithms differ basically from those of the classical machine learning in their manner of treating objects in order to find their hidden structures, besides their performance does not depend on any representations of the objects. 

Our work here concerns supervised and unsupervised binary classification of a set of quantum systems with unknown quantum states based on the concept of quantum fidelity. Quantum fidelity has deep information theoretic roots and connections with differential geometry of Hilbert spaces \cite{book:Nielsen, book:Hayashi, Braunstein, ATR, Alipour1}. The supervised classifier operates on the basis of a training set of quantum systems, assuming that each category is associated with a prior set of quantum systems. We relate the problem of evaluating the category membership to the problem of evaluating fidelity, and propose a method for the latter based on the matrix exponentiation \cite{Lloyd:NP} and quantum phase estimation (\textsc{qpe}) algorithms \cite{book:Nielsen}. By comparing fidelities one can then discern the two classes. In the case of unsupervised classification, we assume we have an oracle which can determine the category membership of the computational basis vectors. This oracle operates as a quantum gate which marks states with one category by shifting their phase, while leaving the rest unchanged. This property allows us to recast the classification problem as a (Grover) quantum search problem, and then again employ the method of evaluating and comparing fidelities to distinguish category memberships. We remark that, interestingly, Grover's search algorithm also can be recast as a finding a geodesic on the manifold of parametric pure quantum states \cite{qinfog-2,qinfog-4}.

The structure of this paper is as follows. In Sec. \ref{sec:fidelity} we lay out a method for evaluating fidelity of two unknown quantum states based on the techniques of density matrix exponentiation and quantum phase estimation. In Secs. \ref{sec:classifier} and \ref{sec:search} we employ the method of evaluating fidelity to construct supervised and unsupervised binary classifiers of quantum systems with unknown quantum states, where in the unsupervised classifier we use Grover's search algorithm. The paper is summarized and concluded in Sec.~\ref{sec:summary}.

\section{Evaluating fidelity}
\label{sec:fidelity}

Quantum fidelity is a measure of similarity or closeness between a pair of quantum states, and thus, it can play a key role in classification of quantum states. In the following, after briefly reviewing the concept of fidelity, we explain a quantum-computational model for evaluating fidelity, which serves the main purpose of this paper better compared to existing methods, e.g., the one in Ref. \cite{fidelity-other}.

Let us assume that we have two quantum systems $A$ and $B$ with the \textit{unknown} density matrices $\varrho$ and $|\sigma\rangle\langle \sigma|$, respectively, both in the Hilbert space $\mathpzc{H}^{\otimes n}$ with $\dim(\mathpzc{H})=2$. Here we show how one can evaluate the fidelity of such states \cite{Nielsen-fidelity}
\begin{equation}
\label{qc1}
\mathpzc{F}(\varrho,\vert\sigma\rangle)= \langle\sigma\vert\varrho\vert\sigma\rangle.
\end{equation}

We assume that we are given a \textsc{qpe} circuit which implements the unitary operation
\begin{equation}
\label{qc3}
U(\varrho,\tau)=e^{i\tau\varrho}= \textstyle{\sum_{j}} e^{2\pi i\theta_j}\vert  v_{j}\rangle\langle  v_{j}\vert,
\end{equation}
where $\varrho=\sum_{j}\lambda_j\vert  v_{j}\rangle\langle  v_{j}\vert$ is the spectral decomposition of $\varrho$, $\theta_j=\lambda_j \tau/(2\pi)$, and $\tau\in(0,1)$ indicates (a predetermined) time. It is also assumed that given any quantum system characterized by an unknown density matrix $\varrho$, the unitary gate $U(\varrho,\tau)$ can be efficiently simulated in order to be utilized within the desired phase-estimation circuit---see Ref. \cite{Lloyd:NP} for the density matrix exponentiation algorithm.

This \textsc{qpe} circuit can be regarded as a quantum black box represented by the unitary operator $ O_{\varrho}^{(\textsc{qpe})}: \mathpzc{H}^{\otimes n}\otimes\mathpzc{H}^{\otimes t}\to\mathpzc{H}^{\otimes n}\otimes\mathpzc{H}^{\otimes t}$, such that for an arbitrary $\vert\Psi\rangle\in\mathpzc{H}^{\otimes n}$, 
\begin{equation}
\label{qc4}
O_{\varrho}^{(\textsc{qpe})}(\vert\Psi\rangle\otimes\vert 0 \ldots 0\rangle)=
\textstyle{\sum_{j=0}^{2^{n}-1}} \Psi_{j}\vert  v_{j}\rangle\otimes\vert\widetilde{\theta}_j\rangle,
\end{equation}
where $\Psi_j=\langle  v_{j}\vert\Psi\rangle$ and $\vert\widetilde{\theta}_j\rangle \equiv |\ell_{j}^{(1)}\rangle \otimes\ldots \otimes |\ell_{j}^{(t)} \rangle$, with $\ell_{j}^{(m)}\in \{0,1\}$ for $\,\forall m\in\{1,\ldots, t \}$, denotes the $t$-qubit estimator of $\theta_{j}\approx\widetilde{\theta}_{j}= 0.\ell^{(1)}_{j}\ldots \ell^{(t)}_{j}$ in the $t$-bit basis \cite{book:Nielsen}. For brevity, hereafter we simply denote $O_{\varrho}^{(\textsc{qpe})}(\vert\Psi\rangle\otimes\vert 0\ldots0\rangle)$ by $ O_{\varrho}^{(\textsc{qpe})}\vert\Psi\rangle$. 

To every state $\vert k\rangle\in\mathpzc{B}_{t}=\{\vert k\rangle\}_{k=0}^{2^{t}-1}$ (the computational basis) we associate a unique real number and a projection as
\begin{align}
\Lambda_k&=(2\pi/\tau)2^{-t}k,\\
P^{(n,t)}_k&=\mathbbmss{I}_{\mathpzc{H}^{\otimes n}}\otimes\vert k\rangle\langle k\vert,
\end{align}
whence we construct the \textsc{qpe} measurement
\begin{align}
F^{(n,t)}&=\textstyle{\sum_{k=0}^{2^{t}-1}} \Lambda_k P^{(n,t)}_{k}.
\end{align}
Assuming that the \textsc{qpe} circuit receives an arbitrary state $\vert\Psi\rangle\in\mathpzc{H}^{\otimes n}$ as the input, by measuring the ``\textsc{qpe}'' observable $F^{(n,t)}$ on the output of the \textsc{qpe} circuit, the average value of the observed outcomes reads as
\begin{align}
\label{qc9}
\mathbbmss{E}_{F}\big[O_{\varrho}^{(\textsc{qpe})}\vert\Psi\rangle\big]&=
\langle\Psi\vert O_{\varrho}^{(\textsc{qpe})}{F^{(n,t)}}O_{\varrho}^{(\textsc{qpe})}\vert\Psi\rangle\nonumber\\
&=\mathpzc{F}(\widetilde{\varrho},|\Psi\rangle),
\end{align}
where $\widetilde{\varrho}=\sum_{j}\widetilde{\lambda}_{j}\vert  v_{j}\rangle\langle  v_{j}\vert$ and $\widetilde{\lambda}_{j}=(2\pi/\tau)\widetilde{\theta}_{j} =\sum_{k=0}^{2^{t}-1}\Lambda_{k}|\langle k\vert\widetilde{\theta}_{j}\rangle|^2$ is indeed an estimation of $\lambda_{j}$ (the average value of the outcomes of the phase-estimation measurement performed on the output state of the \textsc{qpe} circuit, when it receives the eigenvector $\vert  v_{j}\rangle$ as the input). In addition, $\widetilde{\varrho}$ is an estimate or approximation of the state $\varrho$. Equation (\ref{qc9}) implies that the result of this measurement can give an estimate of the fidelity of $\widetilde{\varrho}$ and $|\Psi\rangle$. As a special case, note that $\mathbbmss{E}_{F} \big[O_{\varrho}^{(\textsc{qpe})}\vert  v_{j}\rangle\big] =\widetilde{\lambda}_{j}$.

In the ideal case we have $\{\lambda_j\}_{j=0}^{2^{n}-1}\subseteq \{\Lambda_{k}\}_{k=0}^{2^{t}-1}$, and thus, the \textsc{qpe} is perfect, i.e., $\{\vert\widetilde{\lambda}_{j}\rangle\}_{j=0}^{2^{n}-1}\subseteq\{\vert k\rangle\}_{k=0}^{2^{t}-1}$. In this case,
$\widetilde{\lambda}_j=\lambda_j$; that is, the estimations for the eigenvalues are perfectly accurate. As a result, now Eq. (\ref{qc9}) implies that $\mathbbmss{E}_{F}\big[O_{\varrho}^{(\textsc{qpe})}\vert\sigma\rangle\big]=\mathpzc{F}(\varrho,|\sigma\rangle)$. Thus, in the ideal case, by using $\vert\sigma\rangle$ as the input state, the expected value of the \textsc{qpe} measurement applied on the output state of the circuit, $ O_{\varrho}^{(\textsc{qpe})}\vert\sigma\rangle$, yields the desired quantity (the fidelity of $\varrho$ and $\vert\sigma\rangle$) with perfect accuracy.

In the nonideal case, however, we have $\mathbbmss{E}_{F}\big[O_{\varrho}^{(\textsc{qpe})}\vert\sigma\rangle\big]=
\langle \sigma \vert \widetilde{\varrho}\vert\sigma\rangle$. That is, rather than the desired fidelity ($\mathpzc{F}(\varrho,|\sigma\rangle)$) we obtain an estimate of it ($\mathpzc{F}(\widetilde{\varrho},|\sigma\rangle)$) with the error 
\begin{align}
\label{general6}
\Delta= \left|\mathpzc{F}(\widetilde{\varrho},\vert\sigma\rangle) - {\mathpzc{F}}(\varrho,\vert\sigma\rangle)\right| \leqslant \Vert \varrho -\widetilde{\varrho}\Vert \leqslant \max_{j}|\lambda_{j} - \widetilde{\lambda}_{j}|,
\end{align}
where $\Vert \cdot\Vert$ 
denotes the standard norm 
\cite{book:Nielsen}. Hence to establish an error of magnitude less than $\delta$ in evaluating the fidelity, it is required to reach an approximation of the eigenvalues to $\delta$ accuracy. If $\delta$ is chosen such that $\delta=2^{-m}$, for an integer $m$, then to reach this accuracy with a success probability at least $1-\epsilon$, $t$ is required to satisfy $t=m+\lceil \log(2+1/(2\epsilon))\rceil$ \cite{book:Nielsen}. 

\textit{Remark 1.---}The process of our algorithm has been inspired by the method proposed in Ref. \cite{Lloyd:NP} except for the necessity of a single measurement at our final stage.

\section{Supervised binary classifier}
\label{sec:classifier}

We assume that there is a collection of quantum systems characterized by \textit{unknown} pure states $\{|\sigma_{s}\rangle\}_{s\in I\subset \mathbbmss{N}}\in\mathpzc{H}^{\otimes n}$ each of whose elements belongs to one of the two \textit{known} categories $C_0$ and $C_1$ (note that $C_{0}\cup C_{1} = \{|\sigma_{s}\rangle\}_{s\in I}$ and $C_{0}\cap C_{1}=\varnothing$). We define the function $\Omega: I \to\{0,1\}$ such that $\Omega(s)$ indicates the category to which $|\sigma_{s}\rangle$ belongs. 

\begin{figure}[bp]
\includegraphics[scale=.4]{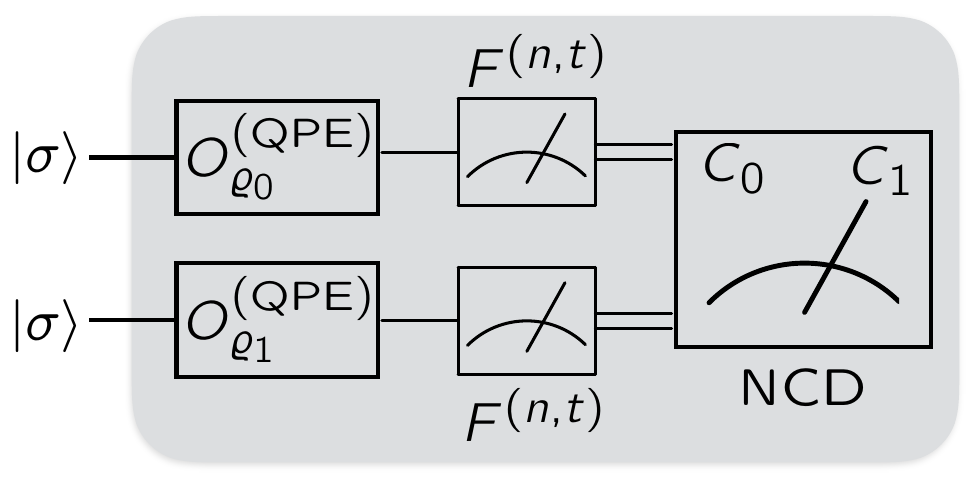}
\caption{Schematic of the supervised binary classifier. Here $O_{\varrho_{a}}^{(\textsc{qpe})}$ indicates the whole
\textsc{qpe} circuit.}
\label{fig:scheme}
\end{figure}

Now we assume that we are provided with two collections $S_0\subseteq C_{0}$ and $S_1\subseteq C_{1}$ of quantum systems as the training subsets and two given probability distributions $\{p_{0}(s)\}_{s\in\Omega^{-1}(0)}$ and $\{p_{1}(s)\}_{s\in\Omega^{-1}(1)}$. These probability distributions can arise from the sampling procedure that generates $S_0$ and $S_1$ and need not to be known. Hence one can assign density matrices 
\begin{align}
\varrho_{a} =\textstyle{\sum_{{s}\in \Omega^{-1}(a)}} p_{a}(s)\vert\sigma_{s}\rangle\langle\sigma_{s}\vert,\,a\in\{0,1\},
\end{align}
to the training subsets. 

We note that the function $\Omega$ is unknown and the goal is indeed to find it or an estimation thereof, $\widetilde{\Omega}$, from the training subsets. This is an instance of supervised machine learning problems, but in the absence of any known representations of objects. We construct the following estimation rule:
\begin{align}
\widetilde{\Omega}(s)=
\begin{cases}
0,~&{\mathpzc{F}}(\varrho_0,\vert\sigma_{s}\rangle)\geqslant{\mathpzc{F}}(\varrho_1,\vert\sigma_{s}\rangle),\cr
1,~&{\mathpzc{F}}(\varrho_0,\vert\sigma_{s}\rangle)<{\mathpzc{F}}(\varrho_1,\vert\sigma_{s}\rangle),
\end{cases}
\end{align}
for any $s\in I$. This function can be obtained by preparing the unitary gates $U(\varrho_{a},\tau)$, for $a\in\{0,1\}$, constructing the \textsc{qpe} circuits $O_{\varrho_a}^{(\textsc{qpe})}$, and next employing the fidelity calculation approach introduced in Sec. \ref{sec:fidelity}, which yield an estimate for the fidelities as
\begin{equation}
\mathbbmss{E}_{F} \big[O_{\varrho_{a}}^{(\textsc{qpe})}\vert\sigma_{s}\rangle\big] =  \mathpzc{F}(\widetilde{\varrho}_{a},\vert\sigma_{s}\rangle),~a\in\{0,1\}.
\end{equation}
Because the evaluation of the fidelities is in general not perfect, rather than $\widetilde{\Omega}(s)$ we obtain an estimation as
\begin{align}
\label{compare}
\widetilde{\Omega}'(s)=
\begin{cases}
0,~&\mathbbmss{E}_{F} \big[O_{\varrho_{0}}^{(\textsc{qpe})}\vert\sigma_{s}\rangle\big]\geqslant\mathbbmss{E}_{F} \big[O_{\varrho_{1}}^{(\textsc{qpe})}\vert\sigma_{s}\rangle\big],\cr
1,~&\mathrm{otherwise},
\end{cases}
\end{align}
which can be discerned by comparing the experimental results with a classical numerical comparing device (\textsc{ncd}). Figure \ref{fig:scheme} illustrates the overall procedure of our supervised binary classifier of quantum systems, which is represented as a blackbox needing two copies of each state of the collection $C_{0}\cup C_{1}$.

\begin{figure}[bp]
\includegraphics[scale=.4]{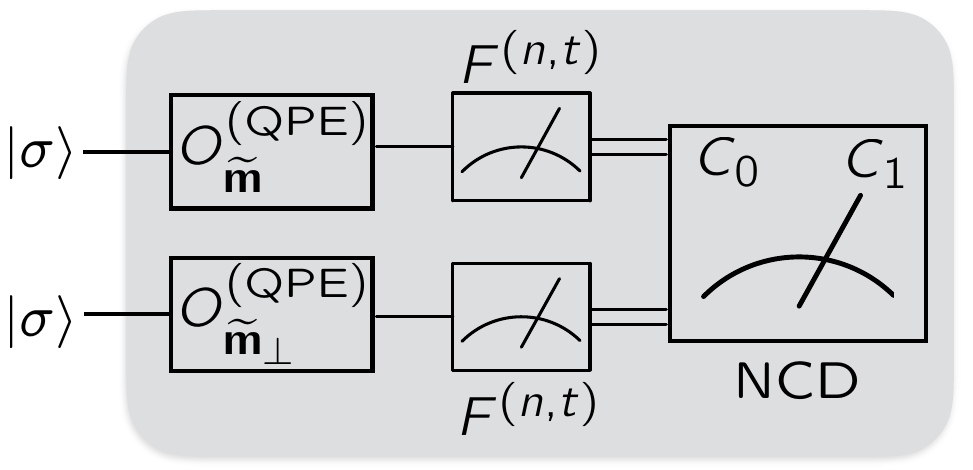}\\
\vskip2mm\hrule\vskip2mm
\includegraphics[scale=.27]{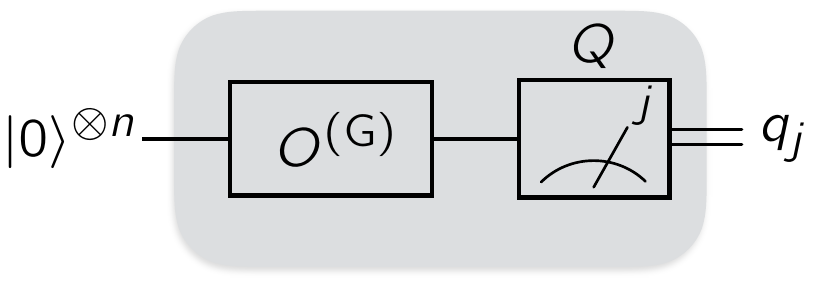}\\
\includegraphics[scale=.27]{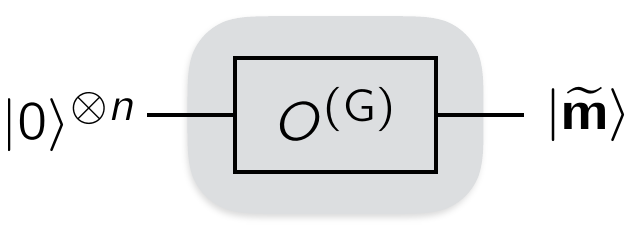} \hskip1mm \includegraphics[scale=.27]{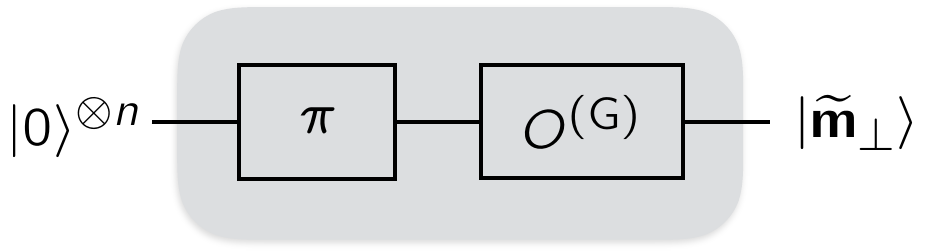}\\
\includegraphics[scale=.25]{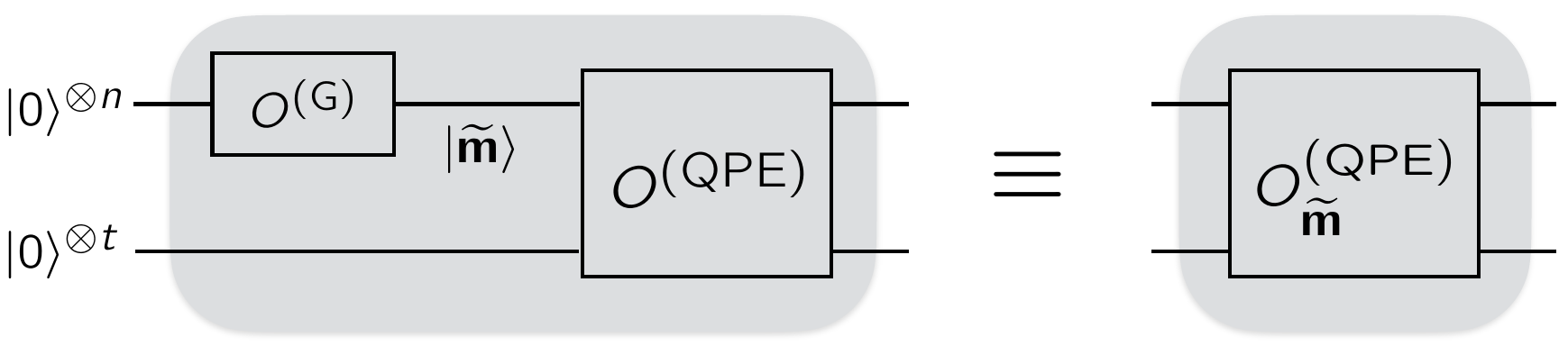} \hskip1mm \includegraphics[scale=.25]{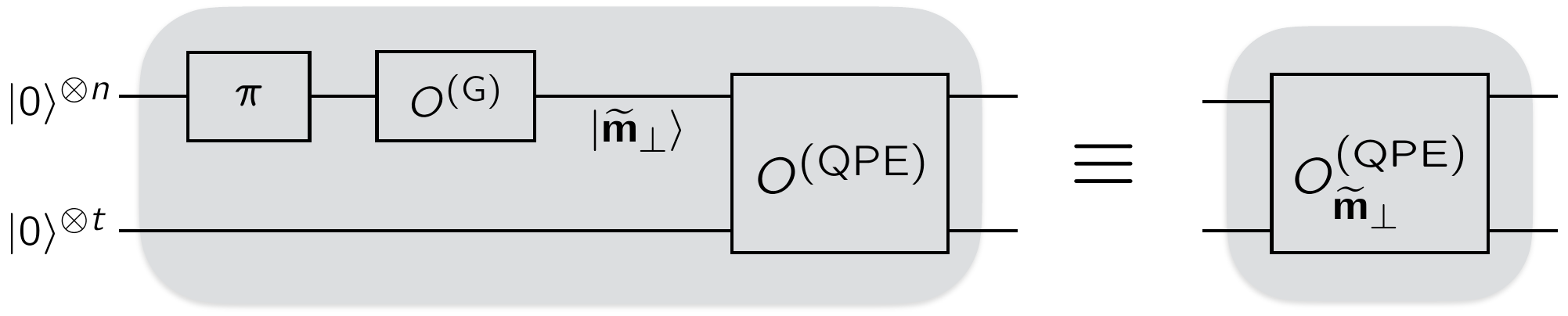}
\caption{Schematic of the unsupervised binary classifier for a general state vector $|\sigma\rangle$ (top). Elements of this classifier are shown below: classifier of the basis vector $|j\rangle$, circuits for generating the states $|\widetilde{\mathbf{m}}\rangle$ and $|\widetilde{\mathbf{m}}_{\perp}\rangle$ (here each box including $O^{(\mathrm{G})}$ refers to the whole Grover's quantum search circuit), and circuits for the operations $O^{(\mathrm{QPE})}_{\widetilde{\mathbf{m}}}$ and $O^{(\mathrm{QPE})}_{\widetilde{\mathbf{m}}_{\perp}}$.}
\label{figure:scheme2}
\end{figure}

\ignore{
\textit{Remark 2.---}In this supervised classifier the \textsc{qpe} circuits are responsible for the run-time. We remind that the complexity of \textsc{qpe} (considering its oracular structure) is $O(t^2)$ \cite{book:Nielsen}.
{\color{red}The overall run-time of this classifier is clearly the same as the run-time of the fidelity evaluation algorithm, $O(r\log d)$.  thus, the complexity of the supervised classifier is the complexity of the phase-estimation circuit, $O(t^2)$.}
}

\section{Unsupervised binary classifier}
\label{sec:search}

Unsupervised machine learning involves classification of objects where there is no subset of the objects having their related categories known \textit{a priori}. Here we demonstrate that the well-known Grover's search algorithm can be considered as an unsupervised binary classifier of quantum states, noting that its performance relies on an oracle which does not require any advance knowledge of the states.

Let us assume that we are given a quantum device which is capable of producing quantum states living in $\mathpzc{H}^{\otimes n}$, in which each state has a unique label from the binary set $\{0,1\}$, corresponding to the classes $C_{0}$ and $C_{1}$. We also assume that we are equipped with an oracle $G$ that can distinguish the labels of the computational basis states. We define the function $\Xi:\mathpzc{H}^{\otimes n}\to\{0,1\}$ such that $\Xi(\vert\sigma\rangle)$ indicates the label of the class to which $\vert\sigma\rangle$ belongs. The action of the oracle can be considered as $O^{(\mathrm{G})}|j\rangle = (-1)^{\Xi(|j\rangle)}|j\rangle$. This oracle is akin to the oracle used in Grover's quantum search algorithm \cite{book:Nielsen}. Thus, we can recast our classification task as a search problem.

We deal with the labels of the basis states in a way that makes these states a collection of objects among which some are the answers to a particular search query while the others are not. We denote by ${B}_0$ the subset of the basis vectors having label ``$0$'' and by ${B}_1$ the remaining subset. Let us introduce
\begin{gather*}
\vert\mathbf{m}\rangle=\frac{1}{\sqrt{M}} \textstyle{\sum_{\vert x\rangle\in{B}_1}}\vert x\rangle,\\
\vert\mathbf{m}_{\perp}\rangle=\frac{1}{\sqrt{N-M}} \textstyle{\sum_{\vert x\rangle\in{B}_0}}\vert x\rangle,
\end{gather*}
where $M=\vert{B}_1\vert$ and $N=2^{n}$. We assume that we have at least two copies of the oracle $O^{(\mathrm{G})}$ to construct two identical quantum search circuits. In addition, we also assume a $\pi$ phase-shift quantum gate such that for any $|\sigma\rangle$, $U_{\pi}\vert\sigma\rangle=-\vert\sigma\rangle$. 

It is straightforward to see that Grover's search algorithm transforms $\vert0\rangle^{\otimes n}$ to $|\widetilde{\mathbf{m}}\rangle$ (where $|\widetilde{\mathbf{m}}\rangle$ is close to $\vert\mathbf{m}\rangle$), whereas the state $\vert0\rangle^{\otimes n}$ is transformed to $|\widetilde{\mathbf{m}}_{\perp}\rangle$ (which is the closest state to $\vert\mathbf{m}_{\perp}\rangle$ from Grover's search channel) if it passes through the $\pi$ phase-shifter before passing through the Grover's search circuit---Fig. \ref{figure:scheme2}. We remind that the oracle $O^{(\mathrm{G})}$ used at the first step of Grover's search algorithm marks the ``answer(s)'' by giving them a phase-shift of $e^{i\pi}$. Hence if an extra $\pi$ phase-shifter precedes $O^{(\mathrm{G})}$, their combination indeed marks the non-answers.

For the special case of the classification of the basis vectors $\{|j\rangle\}$, it suffices to perform the computational-basis measurement $Q:~\{Q_{j}=\vert j\rangle\langle j\vert\}_{j}$ on $|\widetilde{\mathbf{m}}\rangle$ and find the probability  $q_{j}$ of obtaining outcome ``$j$'' ---Fig. \ref{figure:scheme2}. We denote by $\widetilde{\Xi}(\vert j\rangle)$ the assessed label of $\vert j\rangle$, where we set $\widetilde{\Xi}(|j\rangle)= 0$, when $q_{j}=0$; and $\widetilde{\Xi}(|j\rangle)= 1$, otherwise. We, however, remark that imperfections of the search algorithm can affect the reliability of this assessment. 

In the general case, given two copies of an arbitrary state $\vert\sigma\rangle$, one can estimate its class by comparing the fidelities $\mathpzc{F}(\vert\sigma\rangle,|\widetilde{\mathbf{m}}_{\perp}\rangle)$ and $\mathpzc{F}(\vert\sigma\rangle,|\widetilde{\mathbf{m}}\rangle)$.
Since the state $|\widetilde{\mathbf{m}}_{\perp}\rangle$ refers to the class label $0$ and the state $|\widetilde{\mathbf{m}}\rangle$ is a reference to label $1$, it sounds plausible to set the following classification rule:
\begin{align}
\widetilde{\Xi}(\vert\sigma\rangle)=
\begin{cases}
0,~&\mathpzc{F}(\vert\sigma\rangle,|\widetilde{\mathbf{m}}\rangle)\leqslant \mathpzc{F}(\vert\sigma\rangle,|\widetilde{\mathbf{m}}_{\perp}\rangle),\\
1,~&\mathpzc{F}(\vert\sigma\rangle,|\widetilde{\mathbf{m}}\rangle)>\mathpzc{F}(\vert\sigma\rangle,|\widetilde{\mathbf{m}}_{\perp}\rangle).
\end{cases}
\end{align}
Thus, we can employ the fidelity evaluation algorithm here to compute and compare the above fidelities in the same fashion as in Sec. \ref{sec:fidelity}---Fig. \ref{figure:scheme2}.

\ignore{
\textit{Remark 3.---}Similarly to the case of supervised classification, the overall run-time of this classifier is also given by the fidelity evaluation algorithm. The complexity of this classifier is governed by the combined complexity of Grover's search algorithm and the \textsc{qpe} circuit, that is, $O(t^{2}\sqrt{N/M})$.
}

\section{Summary}
\label{sec:summary}

We have developed quantum algorithms based on the well-known methods of density matrix exponentiation and quantum phase estimation to classify quantum systems whose quantum states are unknown, where quantum fidelity have been used as a figure-of-merit to decide category membership. We have also shown how Grover's quantum search algorithm can be considered as an unsupervised method for binary classification of quantum states. Specifically, we have assumed that basis states of a given multiqubit Hilbert space each belong to merely one class out of two specific classes, and that we have also been provided with an oracle whose action is to decide category membership of basis states. Although we have discussed our algorithms for classification of pure states, extensions to classification of mixed states seems straightforward by employing the standard state purification method \cite{book:Nielsen}.

Introducing algorithms of this type can yield a variety of other algorithms which do not require any specific representations of the objects of interest, thus, this approach may enable broader applications in artificial intelligence. We also anticipate that having a quantum-\textsc{ram}, which maps a string of bits to basis vectors of a suitable Hilbert space, our learning methods may also be applied to classify classical data. This can potentially offer a relative speedup over existing classical classification algorithms. One, however, still needs to analyze the tradeoff between complexity and accuracy of such learning algorithms.

\textit{Acknowledgments}.---This work was partially supported by Sharif University of Technology's Office of Vice President for Research and Technology and the Schools of Physics and Nano Science at the Institute for Research in Fundamental Sciences (IPM).


\end{document}